\documentclass[pdflatex,sn-mathphys]{sn-jnl}

\jyear{2022}%

\theoremstyle{thmstyleone}%
\theoremstyle{thmstyletwo}%

\theoremstyle{thmstylethree}%

\begin{document}

\title[Delta Function Scattering with Feynman Diagrams in 1d Quantum Mechanics]{Delta Function Scattering with Feynman Diagrams in 1d Quantum Mechanics}


\author{\fnm{Zakariah} \sur{Crane}}

\affil{\orgaddress{\street{Ne 101st St}, \city{Vancouver}, \postcode{98662}, \state{Washington}, \country{United States}}}


\abstract{In this paper we will demonstrate the use of Feynman Diagrams for one dimensional scattering in quantum mechanics. We will evaluate the S-Matrix explicitly for the Dirac delta and finite wall potentials by summing the full series of Feynman diagrams, illustrating the spirit of perturbation theory. This technique may be useful in introductory quantum mechanics courses, and provides the student with intuition about conservation laws in the context of scattering problems by connecting Feynman diagrams, free propagation, and conservation of the corresponding observable. It also provides a toy model for calculating S-matrix elements in quantum field theory.}




\maketitle

\section{Introduction}\label{sec1}

\ 

When approaching quantum mechanics through the lens of path integrals, it is possible to discuss scattering problems with very similar language to the language which we use to discuss scattering in quantum field theory. This method is useful for the student for its simplicity and for its illustration of what is happening when we write down Green functions and Feynman diagrams.

\ 

When perturbation theory is introduced in introductory material [1,2], it is common to evaluate the terms of order $n$ by throwing out terms above $\lambda^n$. The ability to directly sum the series when presented this way is not very clear, nor is there any physical intuition behind what each term in the series is doing. This method is much more illustrative.

\section{Scattering}\label{sec2}

\ 

It is well known [3] that one can write

\begin{equation} 
\langle \psi_{out} , \psi_{in} \rangle = \int dx \ dt \ \psi_{out}(x_b,t_b) \ G(x_b - x_a, t_b - t_a) \ \psi_{in}(x_a,t_a) 
\end{equation}

\ 

Where, letting $S$ be the action associated with the system,

\[ G(y-x,t-t') = \int Dx' \ e^{iS} \]

\ 

It is also known that one can separate this into a free part and an interacting part by expanding the potential term in $S$ as a power series in its coefficient, which we shall call $\alpha$.

\[ G(x,t) = \int Dx \ e^{i S_0} \left( \sum_{n=0}^{\infty} \frac{(-i \alpha)^n}{n!} \left( \int \ dz \ dt' \ V(z,t')\right)^n \right) \]

\ 

Because we will be summing terms which will contain $n$ integrals over $n$ time variables, it will be necessary for us to account for the ambiguity in the order of integration. As is known, to account for the combinatorics, one must simply multiply the time-ordered integral by a factor of $n!$ [4].

\ 

After examining this on the lattice, one can see that it is permissible for us to separate the free action into a series of free-propagations. The result of this is easily seen to be a series of terms, with the $n$th order term containing $n$ integrations of the potential and $n+1$ free propagators [4]. Additionally, each term in the resulting series can be represented as a graph containing $n+1$ internal lines that connect to $n$ vertices, which is a Feynman diagram.

\

The $0$th order contribution is trivial and can easily be seen to be $F_0 = \delta(p_b - p_a)$. This simply tells us that free propagation conserves momentum. The $1$st order term in the S-matrix serves as an example:

\[ F_1 = -i \alpha \int dx \ dy \int dz \ dt \ \psi_{out}(y,t_b) G_0(y-z,t_b-t) V(z,t) G_0(z-x, t-t_a) \psi_{in}(x,t_a) \]

\ 

Where $G_0$ is the free propagator $G_0 = \int Dx \ e^{i S_0}$. In most cases this series is a challenge to evaluate, and one must be content with only the first few terms. However, in the two cases we will demonstrate, the S-matrix can be evaluated quite easily from these integrals. We will take advantage of the following property of the free propagator.

\[ \int dx \ \psi(y,t_b) G_0(y-z, t_b - t) = \psi(z,t)\] 

\ 

This is useful anywhere that $V=0$, because the corresponding integral reduces to simple free propagation in that case. Now we analyze the case $V(z) = \alpha \delta(z-a)$. Usually when scattering from this potential is discussed, it is necessary to integrate the Schrodinger equation in a small region about $x=a$ in order to acquire the scattering amplitudes. This is not necessary when we use this diagram method. In fact, we will see that scattering from a delta function is the most generic scattering possible.

\ 

Inserting this $V$ into $F_1$, we can quickly evaluate the integral.

\[ F_1 = -i \alpha \int dx \ dy \ \int \ dt \ \psi_{out}(y,t_b) G_0(y-a,t_b - t) G_0(a-x, t-t_b)  \]

\ 

writing $\psi_{out} = \frac{1}{\sqrt{2 \pi}}e^{i(p_b x_b - t_b E_b)}$ and evaluating the integrals, it is easy to see that

\[ F_1 = -i \alpha \ e^{i(p_b - p_a)a}  \int dt \ \frac{e^{i(E_b - E_a)t}}{2\pi} = -i \alpha \ \delta(E_b - E_a) e^{i(p_b - p_a)} \]

\ 

Letting $p = \sqrt{p_b^2} = \sqrt{p_a^2}$ and taking advantage of the fact that

\[ \delta(E_b - E_a) = \frac{m}{p} (\delta(p_b - p_a) + \delta(p_b + p_a)) \]

\ 

It is then easy to see that

\[ F_1 = - i \frac{m \alpha}{p}(\delta(p_b - p_a) + e^{2ip} \delta(p_b + p_a)) \]

\ 

This diagram then contributes both a transmission and a reflection term, while also reflecting the phase change that is picked up due to the reflection.

\ 

Evaluating $F_2$ in a similar way, we find

\[ F_2 = (-i \alpha)^2 \int dz \ dz' \ dt \ dt' \psi_{out}(z,t) \delta(z-a) G_0(z-z',t-t') \delta(z'-a) \psi_{in}(z',t') \]

\ 

\[ F_2 = (-i \alpha)^2 \int dt \ dt' \ \psi_{out}(a,t) G_0(0,t-t') \psi_{in}(z',t') \]
 
\ 

Taking advantage of the substitution under the integral sign $G_0(0,t-t') = \frac{1}{2 \pi \delta(0) } e^{-i(E_b t - E_a t')}$, the final result is

\[ F_2 = (-i \alpha)^2 \ e^{-i(p_b - p_a)} \ \frac{\delta(E_b - E_a)^2}{\delta(0)} = (-\frac{i m\alpha}{p})^2 (\delta(p_b - p_a) + e^{2ip} \delta(p_b + p_a)) \]

\ 

The same thing continues to all orders in perturbation theory.

\[ F_n = (-\frac{i m\alpha}{p})^n (\delta(p_b - p_a) + e^{2ip} \delta(p_b + p_a))\]

\ 

Letting $\Lambda = \frac{m \alpha}{p}$,

\[ S_{ba} = \sum_n F_n = \frac{1}{1-i\Lambda} \delta(p_b - p_a) + e^{2ip} \frac{i \Lambda}{1-i\Lambda} \delta(p_b + p_a) \]

\ 

From which we recover the amplitude. The meaning of our saying "delta scattering is generic" is now clear: the behavior of the scattering is entirely determined by the fact that $V(z)$ does not depend on time. The appearance of these powers of energy conserving delta functions is a simple consequence of the time-independence of the potential, and will always take place.

\ 

This particular potential provides the student with intuition for scattering diagrams, where the spirit of our procedure is clear; We freely propagate the wavefunction to the point(s) where it interacts with the potential, and then we freely propagate it once again to the point of interest.

\

For the finite wall potential, a similar procedure can be followed, with a few modifications. We simply note that in the case of $V(x) = V_0$ inside $[0,a]$, $p_{out} = \sqrt{2mE}$ and $p_{in} = \sqrt{2m(E - V_0)}$. The only modification that occurs when one carries out the calculation is, at first order 

\[ F_1 =  -i V_0 \ \delta(E^{out}_b - E^{out}_a)\int_{0}^{a} dz \ e^{i(p_b - p_a) z} \]

\ 

For the coefficient of $\delta(p_b + p_a)$, one can derive a series which is very similar to the previous case. Each term will be of the form

\[ F_n = \left(-i V_0 \ \delta(E_b - E_a) \ \int_{0}^{a} dz \ e^{i(p_b - p_a)z}\right)^n \]

\ 

The only difference in our amplitude will be that $\Lambda$ changes to become 

\[ \Lambda = \frac{V_0(e^{2ip_a a} - 1)}{4 \sqrt{E(E-V_0)}} \]

\ 

And thereby

\[ S_{ba} = \frac{1}{1 - i \Lambda} \delta(p_b - p_a) + \frac{i \Lambda}{1 - i \Lambda} \delta(p_b + p_a)\]

\ 

Thus, conservation laws can be seen to be a consequence of coordinate-invariance of the corresponding potential, while the generic form of the amplitudes can be understood as arising from energy conservation. The same thing applies to any Hermitian operator, which is especially relevant in 3 dimensions. As long as the potential does not depend on a particular coordinate, then the wavefunction will be separable, and the separable parts will factor out of this integral and leave only the terms which do not commute with the Hamiltonian.

\section{Conclusion}\label{sec13}

\ 

In conclusion, we have demonstrated that it is possible to use this method to solve elementary scattering problems which are of interest to students in their first or perhaps second year of quantum mechanics courses. This method is also immediately applicable to a potential constructed out of any number of delta functions, and there are many other simple potentials for which it is possible to evaluate the full series directly.

\ 

Additionally, this procedure illustrates the concept of free-propagation nicely. It is readily seen that for time-independent scattering, the time-dependent part of the wavefunction freely propagates from $t_b$ to $t_a$. If we were to introduce spin to this particle, and we started it off in a spin eigenstate, we would also clearly see the spin part decouple and freely propagate. In all cases, this procedure makes it very clear that free propagation and conservation laws are intimately related.

\section*{Declarations}

\ 

The author has no competing interests to declare that are relevant to the content of this article. No funding was received to assist with the preparation of this manuscript.

\end{document}